\documentclass[aps,twocolumn,prb,amsmath,amsfonts,showpacs]{revtex4}
\usepackage{graphicx}
\usepackage{comment}
\usepackage{braket}
\usepackage{bm}
\usepackage{txfonts}
\newcommand{\nn}{\nonumber}
\begin{document}
\title{Topological phase transition to Abelian anyon phases due to off-diagonal exchange interaction in the Kitaev spin liquid state }


\author{Daichi Takikawa}
\author{Satoshi Fujimoto}
\affiliation{Department of Materials Engineering Science, Osaka University, Toyonaka 560-8531, Japan }




\date{\today}

\begin{abstract}
We investigate how the Kitaev spin liquid state described by Majorana fermions coupled with $Z_2$ gauge fields is affected by non-Kitaev interactions which exist in real candidate materials of the Kitaev magnet.
It is found that the off-diagonal exchange interaction referred to as the $\Gamma'$  term dramatically changes  the Majorana band structure in the case of the antiferromagnetic Kitaev interaction, and gives rise  to a topological phase transition from a non-Abelian topological phase with
the Chern number equal to $\pm 1$ to an Abelian phase with the Chern number equal to $\pm 2$, in which
 the  $Z_{2}$ vortices behave as Abelian anyons.
On the other hand, other non-Kitaev interactions such as the Heisenberg interaction and the $\Gamma$ term, only affect the bandwidth of the Majorana band as long as the spin liquid state is not destabilized.  
\end{abstract}



\maketitle

\section{Introduction}
Recently, the realization of the Kitaev spin liquid state in real magnetic materials has been extensively explored both theoretically,\cite{kitaev,baskaran2007exact,Jackeli,Rau,PhysRevB.89.235102,Yamaji2014,PhysRevLett.112.207203,PhysRevLett.115.177205,PhysRevB.92.115122,PhysRevLett.119.127204,PhysRevLett.117.037209,udagawa,PhysRevX.8.031032,PhysRevLett.121.147201,PhysRevB.97.241110,Rusna,gordon2019theory,PhysRevResearch.1.033007,minakawa,zhang2019vison} and experimentally.\cite{Liu,Ye,PhysRevB.90.041112,PhysRevB.92.235119,PhysRevB.91.094422,PhysRevB.93.195158,Cao,Baek,Zheng,klanjsek,PhysRevLett.120.117204}
The low-energy excitations in the Kitaev spin liquid state are itinerant Majorana fermions and $Z_2$ vortices (visons), which interact with each other.
In the case with an applied magnetic field, the Majorana fermions acquire a mass gap, and the system is changed into a chiral spin liquid state with chiral Majorana edge states, which is also a non-Abelian topological phase. In this phase, visons with zero-energy Majorana bound states behave as non-Abelian anyons which have implications for the application to quantum computation.\cite{kitaev,RevModPhys.80.1083}
The Kitaev spin liquid is exactly realized in the Kitaev honeycomb-lattice model which is characterized by the Ising-type interactions between nearest-neighbor spins with three different easy-axis directions depending on three types of bonds on the honeycomb lattice.\cite{kitaev} 
In real candidate materials of the Kitaev magnet such as $\alpha$-RuCl$_3$ and Na$_2$IrO$_3$, 
as clarified by {\it  ab initio} studies, 
there are non-Kitaev interactions which do not exist in the ideal Kitaev honeycomb-lattice model, e.g. the Heisenberg exchange interaction, and symmetric off-diagonal exchange interactions referred to as the $\Gamma$ term and the $\Gamma'$ term.\cite{1408.4811,kee,PhysRevB.96.054434,Rau} 
In the case that the non-Kitaev interactions are sufficiently strong compared to the Kitaev interaction, 
the spin liquid state is destabilized, and conventional magnetic orders occur, as extensively discussed in many previous studies.\cite{chaloupka2010kitaev,kimchi2011kitaev,Rau,PhysRevLett.119.157203,Rusna}
However, if the non-Kitaev interactions are not so strong enough to destroy the spin liquid state, 
they may add novel unique features to the Kitaev spin liquid state, which are not expected for the ideal Kitaev model.
In fact, in our previous study,\cite{takikawa} it is found that the off-diagonal exchange interaction $\Gamma'$ term significantly increases the magnitude of the mass gap of itinerant Majorana fermions induced by a magnetic field, and enhances the stability of the topological phase rather than destroy it. 
According to {\it  ab initio} calculations,\cite{kee} the magnitude of $\Gamma'$ in real candidate materials,
 $\alpha$-RuCl$_3$ and Na$_2$IrO$_3$, is rather small compared to that of the Heisenberg interaction and the $\Gamma$ term.
 However, this never implies that $\Gamma'$ is negligible.  
 As a matter of fact, the magnitude of $\Gamma'$ is strongly sensitive to trigonal distortion of the edge-shared tetrahedra structure of these materials.
Thus, for the clarification of sample-dependence of the spin liquid phase of the candidate materials, it is important
to understand effects of the $\Gamma'$ term on the Kitaev spin liquid state.

In this paper, we explore for effects of the non-Kitaev interactions on the spin liquid state more extensively.
The starting point of our argument is the Kitaev spin liquid state perturbed by the above-mentioned non-Kitaev interactions.
We mainly focus on vortex-free states, where the system is described only by itinerant Majorana fermions. 
Our argument is restricted to the parameter regions where the vortex-free phases are not destroyed by the non-Kitaev interactions. 
This assumption is valid as long as the strength of the non-Kitaev interactions are not too large.
We derive an effective Hamiltonian for the itinerant Majorana fermions taking into account the non-Kitaev interactions 
as perturbations, which modify the Majorana band structure.
In general, non-Kitaev interactions may give rise to many-body interactions between Majorana fermions. 
However, unless the coupling strength is large enough, these many-body interactions are irrelevant perturbations to
itinerant Majorana fermions, since the density of states vanishes in the low-energy limit.
Thus, we do not consider any possibility of  spontaneous symmetry breaking due to these interactions, and focus on effects on the band structure of non-interacting Majorana fermions.
One of our main results is that, in the chiral spin liquid phase induced by a magnetic field, the $\Gamma'$-term can give rise to 
a topological phase transition accompanying the change of the first Chern number from $\pm 1$ to $\pm 2$ in the case of the antiferromagnetic Kitaev interaction. 
As elucidated by Kitaev,\cite{kitaev}  the phase with the Chern number $\nu=\pm 2$ is the Abelian topological phase with Abelian anyons $a$, $\bar{a}$, the braiding of which results in the phase change $e^{i\frac{\pi}{4}}$; i.e. they behave as a quarter of a fermion.
On the other hand, it is found that other non-Kitaev interactions such as Heisenberg term and the $\Gamma$-term only change the Majorana band width of the vortex free spin liquid, and does not qualitatively affect the Majorana band structure. 

The organization of this paper is as follows:
In Sec. II, we consider effects of the $\Gamma'$-term, and derive an effective Hamiltonian by using perturbative expansions up
to the second order in $\Gamma'$ around the Kitaev spin liquid state. It is found that the perturbed term change drastically the band structure
of itinerant Majorana fermions.
In Sec. Ill, from numerical analysis of the effective Hamiltonian, 
we demonstrate that topological phase transitions with the change of the Chern number occurs, as the magnitude of $\Gamma'$ increases. 
In Sec. IV, we clarify that the Heisenberg interaction and the $\Gamma$ term do not affect qualitatively the Majorana fermion band.

\section{Effects of the $\Gamma'$ term on the Majorana band structure}
\begin{figure}[http]
 \centering
   \includegraphics[width=7cm]{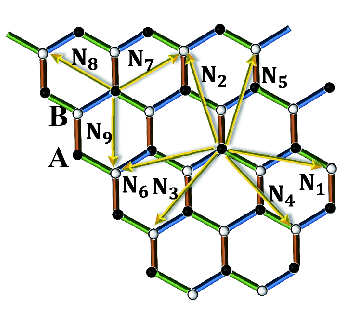}
 \caption{Blue, green, and red edges on the honeycomb lattice represent, respectively, $x$-bonds, $y$-bonds, and $z$-bonds.  Yellow arrows represent the third nearest-neighbor hopping generated by the second order correction term $\mathcal{H}^{(2)}_{\Gamma', eff}$. }
 \label{takikawa1}
\end{figure}

In this section, we investigate effects of the $\Gamma'$ term on the Kitaev spin liquid phase on the basis of
perturbative expansions with respect to $\Gamma'$.
We start with the following Hamiltonian for candidate materials of the Kitaev magnet on a honeycomb lattice such as $\alpha$-RuCl$_3$ and Na$_2$IrO$_3$, 
\begin{eqnarray}
\mathcal{H}_{J}=J\sum_{   \Braket{ij}} \bm{S}_i\cdot\bm{S}_j, \label{eq:hamh}
\end{eqnarray}
\begin{eqnarray}
\mathcal{H}_{K}=-K\sum_{ \Braket{ij}_{\alpha}} S_i^{\alpha}S_j^{\alpha},  \label{eq:hamk}
\end{eqnarray} 
\begin{eqnarray}
\mathcal{H}_{\Gamma}=\Gamma\sum_{\substack{    \Braket{ij}_{\alpha} \\ \beta,\gamma \neq \alpha}} [S_i^{\beta}S_j^{\gamma}+S_i^{\gamma}S_j^{\beta}], \label{eq:hamgam1}
\end{eqnarray}
\begin{eqnarray}
\mathcal{H}_{\Gamma'}=\Gamma'\sum_{\substack{  \Braket{ij}_{\alpha} \\ \beta \neq \alpha}} [S_i^{\alpha}S_j^{\beta}+S_i^{\beta}S_j^{\alpha}],  \label{eq:hamgam2}
\end{eqnarray}
where $S^{\alpha}_i$ is an $\alpha=x,y,z$ component of an $s=1/2$ spin operator at a site $i$.
 $\mathcal{H}_{J}$ is the Heisenberg exchange interaction between the nearest neighbor sites, and
$\mathcal{H}_{K}$ is the Kitaev interaction.
Here, $\Braket{ij}_{\alpha}$ denotes that the $i$-site and the $j$-site are the nearest-neighbor sites connected via an $\alpha$-bond on the honeycomb lattice
(see FIG.\ref{takikawa1}).
$\mathcal{H}_{\Gamma}$ and $\mathcal{H}_{\Gamma'}$ are symmetric off-diagonal exchange interactions arising from spin-orbit couplings and oxygen-mediated exchange interactions in the edge-shared octahedra structure.\cite{Rau}\\

The ideal Kitaev Hamiltonian $\mathcal{H}_K$ is exactly solvable in terms of the Majorana fermion representation:
\begin{eqnarray}
S^x_j=\frac{\rm i}{2}b^x_jc_j, ~S^y_j=\frac{\rm i}{2}b^y_jc_j, ~S^z_j=\frac{\rm i}{2}b^z_jc_j,
\end{eqnarray}
where $b_j^{\alpha}$ ($\alpha=x,y,z$) and $c_j$ are Majorana fermion operators, and the Hilbert space where these operators act on is restricted 
to satisfy $ D_i|\phi\rangle=|\phi\rangle$ 
with $D_i=b_i^xb_i^yb_i^zc_i$ and $|\phi\rangle$ the eigen state of the Kitaev spin liquid.
In terms of the Majorana fields, $\mathcal{H}_K$ is expressed as,
\begin{eqnarray}
\mathcal{H}_K=\frac{{\rm i}}{4}\sum_{i,j}\hat{A}_{ij}c_ic_j,   \label{eq:HK-ma}
\end{eqnarray}
where $\hat{A}_{ij}=\frac{1}{2}K\hat{u}_{ij}$ and $\hat{u}_{ij}={\rm i}b_i^{\alpha}b_j^{\alpha}$  with $i,j \in \alpha$-bond.
The $Z_2$ gauge fields $\hat{u}_{ij}$ commute with $\mathcal{H}_K$, and can be replaced by the eigenvalues $\pm 1$. 
For $K>0$ ($K<0$), in the ground state, we can put $\hat{u}_{ij} \rightarrow 1$ ($-1$), and hence, $\hat{A}_{ij}\rightarrow \frac{1}{2}K$ ($-\frac{1}{2}K$).
Then, eq. (\ref{eq:HK-ma}) is reduced to the Hamiltonian of free massless Majorana fermions which can be diagonalized in the momentum representation. 
When the sign of a $Z_2$ gauge field is flipped, a $Z_2$ vortex (vison), the excitation energy of which is $\sim K$, is created.\cite{kitaev}

\begin{figure}[http]
 \centering
   \includegraphics[width=6cm]{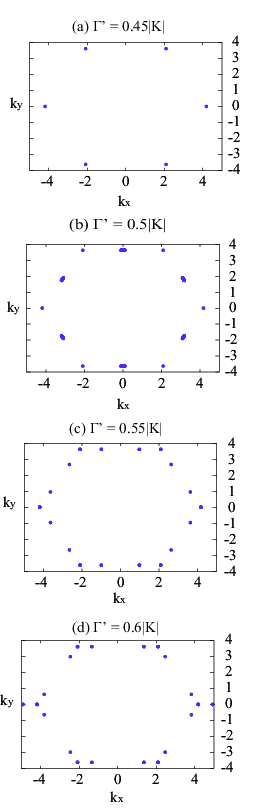}
 \caption{Dirac points in the Brillouin zone for several values of $\Gamma^{\prime}$ in the case of the antiferromagnetic Kitaev interaction.}
 \label{takikawa2}
\end{figure}

In this paper, we consider effects of non-Kitaev interactions, $\mathcal{H}_J$, $\mathcal{H}_{\Gamma}$, and $\mathcal{H}_{\Gamma'}$, on the vortex-free spin liquid phase which is the ground state of $\mathcal{H}_K$.
As a first step, in this section, we focus on the $\Gamma'$-term, which, as shown below, drastically affects the band structure of Majorana fermions. 
To see effects on the Kitaev spin liquid state, 
following the spirit of the original Kitaev's paper,\cite{kitaev} we carry out perturbative expansions with respect to $\Gamma'$
around the vortex-free Kitaev spin liquid state, which is separated from excited states with visons by a finite energy gap $\sim K$.
The corrections to the ground state of the ideal Kitaev Hamiltonian $\mathcal{H}_K$ are expressed in terms of perturbative expansions of the self-energy due to $\mathcal{H}_{\Gamma'}$,
\begin{eqnarray}
&\Sigma(E)=\Pi_{0}(\mathcal{H}_{\Gamma'}+\mathcal{H}_{\Gamma'}G'_{0}(E)\mathcal{H}_{\Gamma'}+...)\Pi_{0} \\
&G'_{0}(E_{0}) \sim -\frac{1-\Pi_{0}}{|K|}
\end{eqnarray}
where, $\Pi_{0}$ is a projection to the vortex-free spin liquid state.
Up to the second order in $\mathcal{H}_{\Gamma'}$, we obtain the perturbative corrections to the effective Hamiltonian,
\begin{eqnarray}
\mathcal{H}^{(1)}_{\Gamma', eff}&=&0, \\
\mathcal{H}^{(2)}_{\Gamma', eff}&=&\Pi_{0}(\mathcal{H}_{\Gamma'}G'_{0}(E)\mathcal{H}_{\Gamma'})\Pi_{0} \nn \\ 
&\sim &- \frac{1}{|K|}\Pi_{0} \mathcal{H}_{\Gamma'}\mathcal{H}_{\Gamma'}\Pi_{0}\nn \\ 
&= &-\frac{{\Gamma^{\prime}}^{2}}{16|K|}
 {\displaystyle \sum_{\begin{subarray}{c}   \Braket{ij}_{\alpha} \\ \bf{\chi \neq \alpha} \\ \bf{\tau \neq \chi, \alpha}
 \\ \end{subarray}}
 \sum_{\begin{subarray}{c}  \Braket{lk}_{\xi} \\ {v \neq \xi} \\ {w\neq v , \xi } \\ \end{subarray}}}   
 S^{\alpha}_{i}S^{\chi}_{j}S^{\xi}_{l}S^{v}_{k},\label{eq:second order}
\end{eqnarray}
where $\alpha,\chi,\xi,\nu = x,y,z$.
In eq.(\ref{eq:second order}), the second order term arises from  the off-diagonal exchange interactions acting on the $\alpha$-bond, and the $\xi$-bond.
To analyze this term more precisely, we use the fact that in the Majorana fermion representation of the Kitaev spin liquid state, 
gauge Majorana fields $b_i^{\alpha}$ should be paired on the $\alpha$-bond connecting two sites $i$ and $j$ to form $Z_2$ gauge fields $\hat{u}_{ij}^{\alpha}={\rm i}b^{\alpha}_ib^{\alpha}_j$, since the Kitaev spin liquid state is expressed by the eigen state of the $Z_2$ gauge fields.
Then, eq.(\ref{eq:second order}) is recast into,
\begin{eqnarray}
\mathcal{H}^{(2)}_{\Gamma',eff}&=-\frac{{\Gamma^{\prime}}^{2}}{16|K|}\Biggl[
{\displaystyle \sum_{\begin{subarray}{c} \Braket{ij}_{\alpha} \\ \Braket{jk}_{\tau}  \\ \Braket{kl}_{\alpha} \\ \bf{\tau \neq \chi,\alpha}
 \\ \end{subarray}}}
   ic_{l}c_{i}\hat{u}^{\alpha}_{ij}\hat{u}^{\tau}_{kj}\hat{u}^{\alpha}_{kl}\nn\\
&-{\displaystyle \sum_{\begin{subarray}{c} \Braket{ij}_{\alpha} \\ \Braket{jk}_{\tau}  \\ \Braket{kl}_{\chi} \\ \bf{\tau \neq \chi, \alpha}
 \\ \end{subarray}}}
   ic_{l}c_{i}\hat{u}^{\alpha}_{ij}\hat{u}^{\tau}_{kj}\hat{u}^{\chi}_{kl}
\qquad\Biggr]. \label{eq:gammagammagamma}
\end{eqnarray}
Therefore,  for the vortex-free spin liquid state, we have,
\begin{eqnarray}
\mathcal{H}^{(2)}_{\Gamma', eff}=& -\frac{{i\Gamma^{\prime}}^{2}}{16|K|}
 {\displaystyle \sum_{ \bf{i }}}\Biggl[
{\displaystyle \sum_{\begin{subarray}{c} {p=1\sim 6 } \\ {m=0,1} 
 \\ \end{subarray}}}
   (-1)^{m}c_{i+N_{p}(-1)^{m-1}}c_{i}\nn\\
   &+2{\displaystyle \sum_{\begin{subarray}{c} {p=7,8,9 } \\ {m=0,1} 
 \\ \end{subarray}}}
   (-1)^{m-1}c_{i+N_{p}(-1)^{m-1}}c_{i},
\Biggr].\label{eq:majorana gamma}
\end{eqnarray}
where $N_{p}$ is a vector defined in FIG.\ref{takikawa1}, and  $m = 0$ if the site $i$ is on the $A$ sub-lattice of the honeycomb
lattice, and $m = 1$ if the site $i$ is on the $B$ sub-lattice. The factor $2$ in front of the second term arises from two shortest paths connecting sites
$i$ and $i\pm N_p$ with $p=7,8,9$.
This second-order perturbation term generates the third nearest-neighbor hopping of itinerant Majorana fermions as shown in FIG.\ref{takikawa1},
and changes the Majorana band structure drastically.
The total effective Hamiltonian, $\mathcal{H}_{eff}=\mathcal{H}_K+\mathcal{H}^{(2)}_{\Gamma',eff}$, is,
\begin{eqnarray}
\mathcal{H}_{eff}=\sum_{\bm{k}}
(c_A(-\bm{k}),c_B(-\bm{k}))\left(
\begin{array}{cc}
0 & i  F(\bm{k}) \\
 -i  F^{\ast} (\bm{k})  & 0
\end{array}
\right)
\left(
\begin{array}{c}
c_A(\bm{k}) \\
c_B(\bm{k})
\end{array}
\right)
\label{eq:fourierKitaev}
\end{eqnarray}
with $F(\bm{k})=f(\bm{k})+g(\bm{k})$, and
\begin{eqnarray}
f(\bm{k})=\frac{K}{2}(e^{{\rm i}\bm{k}\cdot\bm{n}_1}+e^{-{\rm i}\bm{k}\cdot\bm{n}_2}+1),
\end{eqnarray}
\begin{eqnarray}
g(\bm{k})&= \frac{{\Gamma^{\prime}}^{2}}{4|K|}
 {\displaystyle \sum_{\begin{subarray}{c} P=1\sim 6 \\  \end{subarray}}}
   \mathrm { e } ^ { \mathrm {- i } \bm{k} \cdot \bm{ N}_{P}}
   -\frac{{\Gamma^{\prime}}^{2}}{2|K|}{\displaystyle \sum_{\begin{subarray}{c} P=7,8,9   
 \\ \end{subarray}}}
   \mathrm { e } ^ { \mathrm { -i } \bm {k} \cdot \bm{ N}_{P}}\nn\\
  & +\frac{K}{2}\bigl[1+   \mathrm { e } ^ { \mathrm { i } \bm{k} \cdot \bm{ n}_{1}}+   \mathrm { e } ^ { \mathrm { -i } \bm{k} \cdot \bm{ n}_{2}}\bigr], \label{eq:gamma-secondorder}
\end{eqnarray}
where  $c_{A(B)}(\bm{k})$ is a Majorana field on the $A$ ($B$) sub-lattice of the honeycomb lattice, and
$\bm{n}_{1} = ( \frac{1}{2} ,\frac{\sqrt{3}}{2})$, $\bm{n}_{2} = ( \frac{1}{2} ,-\frac{\sqrt{3}}{2})$.
The term $g(\bm{k})$ arises from $\mathcal{H}_{\Gamma', eff}^{(2)}$.
In FIG.\ref{takikawa2},
we plot zero-energy Dirac points of the Majorana band of eq.(\ref{eq:fourierKitaev}), i.e. $|F(\bm{k})|=0$, for several values of $\Gamma'/|K|$.
We see that the second-order perturbation term changes the location and the number of Dirac points in the Brillouin zone, which implies
the change of the Chern number in the case with a mass gap of Majorana fermions induced by an applied magnetic field.
We explore for this possibility in the next section.

\section{Topological phase transition with the change of the Chern number and Abelian anyon phase}
In this section, on the basis of the results obtained in Sec. II, we discuss topological phase transitions induced by the second-order perturbation term
$\mathcal{H}^{(2)}_{eff}$
in the case with a mass gap of itinerant Majorana fermions which is generated by an applied magnetic field.
For the ideal Kitaev model, eq.\eqref{eq:hamk}, when time-reversal symmetry is broken by an applied magnetic field,
itinerant Majorana fermions acquire a mass gap and a non-Abelian topological phase with the Chern number equal to $\pm 1$ is realized.
This topological feature is changed by $\mathcal{H}^{(2)}_{eff}$ as seen below.
 In the case with a magnetic field $\bm{h}=(h_x,h_y,h_z)$, we obtain
 the effective Hamiltonian $\mathcal{H}_{eff, h}$ from perturbative calculations up to the third order in $\bm{h}$, 
\begin{eqnarray}
&\mathcal{H}_{eff,h}&=\mathcal{H}_{K}+\mathcal{H}^{(2)}_{\Gamma^{\prime}, eff}+\mathcal{H}^{(2)}_{h,\Gamma^{\prime}}+\mathcal{H}^{(3)}_{h}\label{eq:allham},
\end{eqnarray}
where the third and fourth terms in the right-hand side
are,
\begin{eqnarray}
&\mathcal{H}^{(2)}_{h,\Gamma^{\prime}}&=-{\rm i}\frac{\Gamma'}{4|K|}\sum_{p,m}\sum_{\substack{   \Braket{ij}_{\alpha} \\  \Braket{jk}_{\beta} \\ i=k+\bm{n}_p}}(h_{\alpha}+h_{\beta})(-1)^m
c_ic_k, \label{eq:pert-gap}   \\
&\mathcal{H}^{(3)}_{h}&= {\rm i}\frac{h_xh_yh_z}{K^2}\sum_{i,m}\sum_{p=1,2,3}(-1)^m c_{i+\bm{n}_p}c_i. \label{eq:mass1}
\end{eqnarray}
Here, 
$\bm{n}_{3} = -\bm{n}_{1}-\bm{n}_{2}$, and $m = 0$ if the site $i$ is on the $A$ sub-lattice of the honeycomb lattice, 
and $m = 1$ if the site $i$ is on the $B$ sub-lattice.
These term generates the Majorana mass term,
\begin{eqnarray}
\Delta(\bm{k})&=&\Delta_0(\bm{k})+\Delta_1(\bm{k}), \\
\Delta_0(\bm k)&=&\frac{4h_{x}h_{y}h_{z}}{|K|^{2}}\bigl[\sin(\bm{k} \cdot \bm{n}_{1})+\sin(\bm{k} \cdot \bm{n}_{2})\nn\\
&+&\sin(\bm{k} \cdot \bm{n}_{3})\bigr], \\
\Delta_1(\bm k)&=&-\frac{\Gamma^{\prime}}{|K|}\bigl[
(h_{x}+h_{z})\sin(\bm{k} \cdot \bm{n}_{1})+(h_{y}+h_{z})\sin(\bm{k} \cdot \bm{n}_{2})\nn\\
&+&(h_{x}+h_{y})\sin(\bm{k} \cdot \bm{n}_{3})\bigr]. \nn\\
\end{eqnarray}
The mass gap terms change the system into topological phases
with the nonzero Chern number.
Note that the mass term eq.(\ref{eq:pert-gap}) linear in $\bm{h}$ was obtained before in ref.\cite{takikawa}.

\begin{figure}[http]
 \centering
   \includegraphics[width=8cm]{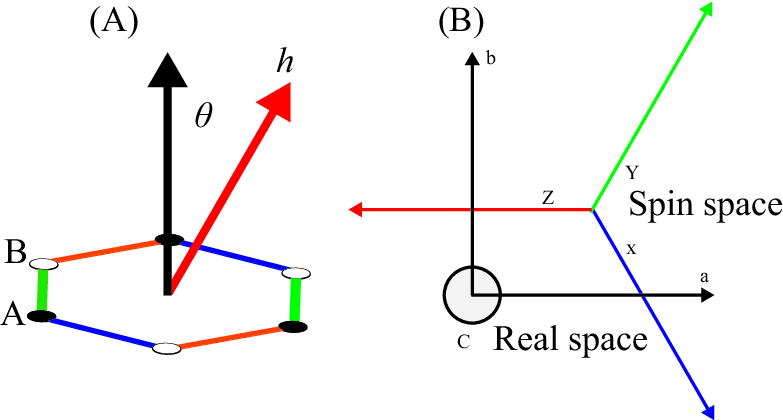}
 \caption{(A) The direction of an applied magnetic field tilted from the vertical direction to the ab-plane by angle $\theta$. (B) Spin axes of the Kitaev magnet $\alpha$-Rucl$_3$}
 \label{takikawa3}
\end{figure}

\begin{figure}[http]
 \centering
   \includegraphics[width=6cm]{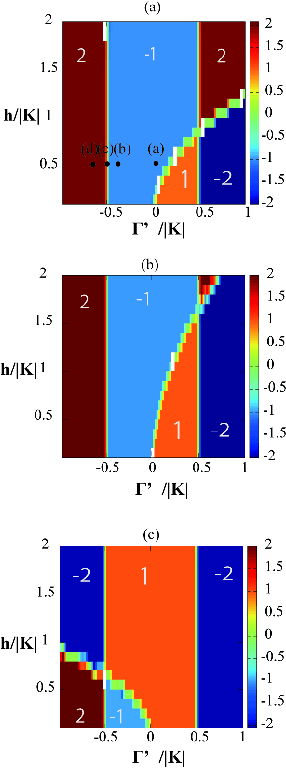}
 \caption{The first Chern number plotted as a function of $\Gamma^{\prime}$ and a magnetic field $h$ for several values of the field angle $\theta$ in the case of the  antiferromagnetic Kitaev interaction. (a) $\theta=0$, (b) $\theta=\pi/6$, (c) $\theta=\pi/3$}
 \label{takikawa4}
\end{figure}

\begin{figure}[http]
 \centering
   \includegraphics[width=6cm]{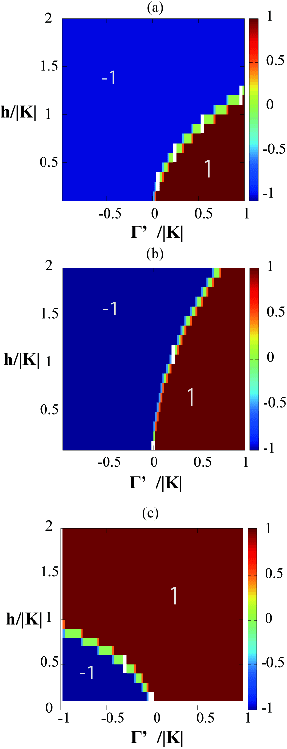}
 \caption{The first Chern number plotted as a function of $\Gamma^{\prime}$ and a magnetic field $h$ for several values of the field angle $\theta$ in the case of the ferromagnetic Kitaev interaction. (a) $\theta=0$, (b) $\theta=\pi/6$, (c) $\theta=\pi/3$}
 \label{takikawa5}
\end{figure}

To investigate topological characters of this gapped phase,
we calculate the first Chern number $\nu$ which is given by,
\begin{eqnarray}
\nu&=&\frac{1}{2\pi}\int_{Bz} dk_xdk_y \ \Omega_{k_xk_y},\\
\Omega_{k_xk_y}&=&\frac{1}{2}\ \hat{\mathcal{H}}\cdot (\partial_{k_x}\hat{\mathcal{H}} \times \partial_{k_y}\hat{\mathcal{H}}), \ \hat{\mathcal{H}}:=\frac{\mathcal{H}}{|\mathcal{H}|},   \\
\nn
\end{eqnarray}
with $\Omega_{k_xk_y}$ the Berry curvature of the Majorana band.
We consider the case with a magnetic field applied in the direction shown in FIG.\ref{takikawa3}(A).
The calculated results of the Chern number as a function of $h=|\bm{h}|$ and $\Gamma'$ are plotted in FIGs.\ref{takikawa4} and \ref{takikawa5}.
We see that,  in the case of the antiferromagnetic Kitaev interaction $K<0$, as shown in FIG.\ref{takikawa4}, 
topological phase transitions with the change of the Chern number from $\pm 1$ to $\pm 2$ occur,
as the magnitude of $\Gamma'$ increases.
This topological phase transition may be experimentally observed by the measurement of the quantized thermal Hall effect.\cite{kasahara2018unusual,kasahara,yokoi}
The changes of the Chern number are in accordance with the change of the number of Dirac cones 
shown in FIG.\ref{takikawa2}. 
On the other hand, in the case of the ferromagnetic Kitaev interaction $K>0$,
the phase with the Chern number equal to $\pm 2$ is not realized for any values of $\Gamma'$,
as shown in FIG.\ref{takikawa5}.

Another interesting feature found in FIGs.\ref{takikawa4} and \ref{takikawa5} is that the sign of the Chern number, i.e. the sign of the thermal Hall conductivity, is flipped from $+1$ to $-1$ (or $-1$ to $+1$) depending on the sign of $\Gamma'$. This phenomenon occurs for both the aintiferromagnetic and ferromagnetic Kitaev interactions.

\begin{figure}[http]
 \centering
   \includegraphics[width=6cm]{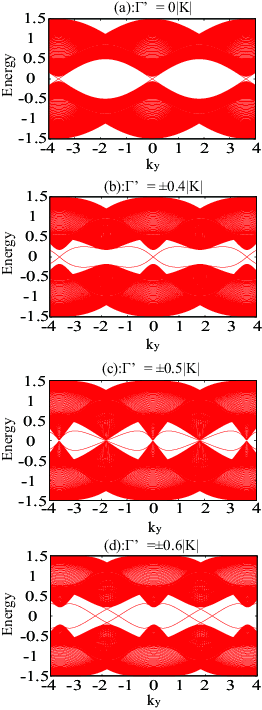}
 \caption{
 The energy spectra in the case with armchair open boundaries and the antiferromagnetic Kitaev interaction.
  A magentic field is set to $h=0.5|K|$ and $\theta=\pi/6$. 
  The parameters correspond to the points (a), (b), (c), and (d) denoted in FIG.\ref{takikawa4}(a). (a) $\nu=-1$ with $\Gamma^{\prime}=0|K|$, 
  (b) $\nu=-1$ with $\Gamma^{\prime}=0.4|K|$, (c) the topological phase transition point at $\Gamma^{\prime}=0.5|K|$, (d)  $\nu=2$ with $\Gamma^{\prime}=0.6|K|$. There is one edge state for (a) and (b), while there are two for (d).}
 \label{takikawa6}
\end{figure}

We also examine the bulk-edge correspondence by calculating edge states in the case with open boundaries. The calculated energy spectra 
 in the antiferromagnetic Kitaev interaction are shown in FIG.\ref{takikawa6}. 
Here, we consider the system with armchair open edges to avoid extrinsic complexity caused by 
zigzag boundaries 
which lead to non-topological flat bands of the edge states unrelated to the bulk Chern number.\cite{doi:10.1143/JPSJ.65.1920}
In this calculation, the total number of the unit cell along the $a$-axis is $100$, a magnetic field is set to $h=0.5|K|$, and the field direction is $\theta=\pi/6$.
The results are shown in FIG.\ref{takikawa4}.
We see that there are two chiral edge states in the case of the Chern number $\nu=\pm 2$ (see FIG.\ref{takikawa4}(d)).

The emergence of the topological phase with the even Chern number is quite intriguing, because, in this case, Abelian topological phases with
Abelian anyons are realized.
As elucidated by Kitaev\cite{kitaev}, when $\nu=2$ (mod $4$), there are two types of the $Z_2$ vortices which behave as Abelian anyons $a$, $\bar{a}$. 
They obey the fusion rule
 $a\times a=\bar{a}\times\bar{a}=\varepsilon$, $a\times\bar{a}=1$, $a\times \varepsilon =\bar{a}$, and $\bar{a}\times \varepsilon =a$,
 where $\varepsilon$ is a fermion field.
 The braiding of $a$ and $\bar{a}$
accompanies the phase change $e^{i\frac{\pi}{4}}$, which implies that $a$ and $\bar{a}$ are  neither bosons nor fermions, but
behave as a quarter of a fermion. 
It is an interesting future issue to explore for novel phenomena associated with these Abelian anyons in Kitaev magnets.

\section{Effect of other non-Kitaev interactions}

In this section, we discuss effects of  other non-Kitaev interactions, i.e. 
the Heisenberg term eq.\eqref{eq:hamh} and
the off-diagonal exchange interaction $\Gamma$ term eq.\eqref{eq:hamgam1} on the vortex-free spin liquid state.
We do not discuss the instability of the Kitaev spin liquid state toward conventional magnetic ordered states induced by these interactions,
which has been discussed in many previous studies.\cite{chaloupka2010kitaev,kimchi2011kitaev,Katukuri_2014,PhysRevLett.119.157203,jiang2019field} 
However, instead, we consider the issue how properties of the spin liquid state are affected by these interactions provided that
 the ground state is described by Majorana fermions coupled with $Z_2$ gauge fields.
  We deal with these non-Kitaev interactions as  perturbations to the Kitaev spin liquid state.
 The main result of this section is that both the $\Gamma$ term and the Heisenberg term generate nearest-neighbor hopping terms of itinerant Majorana fermions, which only affect the Majorana band width,  and do not give rise to qualitative changes of the Majorana band structure.

\begin{figure}[http]
 \centering
   \includegraphics[width=5cm]{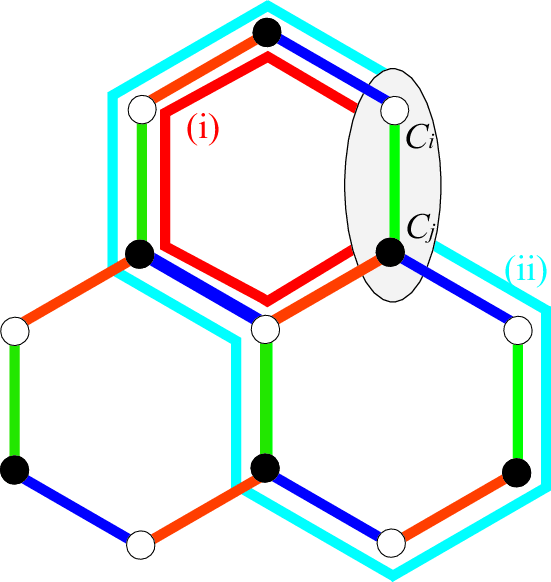}
 \caption{The nearest neighbour hopping term between sites $i$ and $j$ generated from the off-diagonal exchange interaction $\Gamma$ term and the Heisenberg interaction term. The path (i) corresponds to the lowest order perturbation term which includes $\Gamma S_i^yS_j^x$. The path (ii) corresponds to the lowest order term which includes $JS_i^yS_j^y$.}
 \label{takikawa7}
\end{figure}

Since the Heisenberg term 
 $JS^{\alpha}_iS^{\alpha}_j$ acting on $\langle ij\rangle_{\alpha}$ sites trivially normalizes the Kitaev interaction, we focus on other terms.
Putting $V^{\prime}=\mathcal{H}_{J}'+\mathcal{H}_{\Gamma}$,
where $\mathcal{H}_J'$ is the Heisenberg interaction term in which all terms $JS^{\alpha}_iS^{\alpha}_j$ acting on $\langle ij\rangle_{\alpha}$ sites are eliminated,
we carry out the perturbative calculation with respect to $V'$.
The $n$-th order perturbation is given by,
\begin{eqnarray}
\mathcal{H}^{(n)}_{eff}&=&\Pi_{0}V^{\prime}G'_{0}(E)V^{\prime}G'_{0}(E)V^{\prime}\cdots V^{\prime}\Pi_{0}\nn\\
&\sim &\frac{(-1)^{n-1}}{|K|^{n-1}}\Pi_{0}V^{\prime}(1-\Pi_{0})V^{\prime}(1-\Pi_{0})V^{\prime}\cdots V^{\prime}\Pi_{0}\nn\\
&=&\sum^{n}_{m=0}
\Biggl(
\begin{array}{cc}
n \\
m  
\end{array}\Biggr)
\Pi_{0}(\mathcal{H}_{J})^{n-m}(\mathcal{H}_{\Gamma})^{m}\Pi_{0}\nn\\
 &=&\sum^{n}_{m=0}\frac{(-1)^{n-m}  J^{n-m}\Gamma^{m} }{4^{n}}
\Biggl(
\begin{array}{cc}
n \\
m  
\end{array}\Biggr)
\mathcal{F}^{(n,m)}(\sigma_{i},\sigma_{j}\cdots),
\label{nthorder}
\end{eqnarray}
where $\mathcal{F}^{(n,m)}(\sigma_{i},\sigma_{j}\cdots)$ is defined as,
\begin{eqnarray}
\mathcal{F}^{(n,m)}(\sigma_{i},\sigma_{j}\cdots)
 \coloneqq \Pi_{0}
 \Bigg[\sum_{\substack{  \Braket{ij}_{\alpha} \\ \beta,\gamma \neq \alpha}} {\sigma}_i^{\beta}{\sigma}_j^{\gamma}+{\sigma}_i^{\gamma}{\sigma}_j^{\beta}  \Biggr]^{m} \times \nn\\
 \Biggl[ \sum_{  \Braket{ij}_{\alpha} } \sigma^{\beta}_{i}\sigma^{\beta}_{j}+\sigma^{\gamma}_{i}\sigma^{\gamma}_{j} \Biggr] ^{n-m}\Pi_{0}.
\end{eqnarray}
As mentioned above, we focus on perturbation terms which are expressed in the quadratic form of itinerant Majorana fields $c_i$ in the vortex-free ground state.
Let us consider the case that one of Majorana fields in a quadratic term, $c_i$, arises from $\Gamma S_i^{\beta}S_j^{\gamma}=-\frac{\Gamma}{4}b_i^{\beta}c_ib_j^{\gamma}c_j$ acting on $\langle i j \rangle_{\alpha}$ where $\alpha,\beta,\gamma$ are given by the cyclic permutation of $x$, $y$ , $z$.
Then, since,
\begin{eqnarray}
-\frac{\Gamma}{4}b_i^{\beta}c_ib_j^{\gamma}c_j|\phi\rangle=\frac{\Gamma}{4}b_i^{\beta}c_ib_j^{\alpha}b_j^{\beta}|\phi\rangle,
\end{eqnarray}
there are two cases; (i) another itinerant Majorana field in the quadratic term is $c_j$, and $b^{\beta}_i$ and $b^{\gamma}_j$ are formed into $Z_2$ gauge fields on the $\beta$-bond and the $\gamma$-bond, respectively. (ii) $b^{\alpha}_j$ and $b^{\beta}_j$ are formed into $Z_2$ gauge fields
on the $\alpha$-bond and $\beta$-bond, respectively. However, the second case (ii) is not possible, because the $Z_2$ gauge field on the $\alpha$-bond, ${\rm i}b^{\alpha}_ib^{\alpha}_j$, can not be formed with the lack of $b^{\alpha}_i$.
Thus, only the nearest-neighbor hopping term of itinerant Majorana fermions $c_ic_j$ can be realized.
This argument is also applicable to the case that $c_i$ arises from the Heisenberg interaction $JS_i^{\beta}S_j^{\beta}$ or $JS_i^{\gamma}S_j^{\gamma}$.  
On the basis of this insight, it is found that 
all perturbation terms which lead to nearest-neighbor hopping between sites $i$ and $j$
are expressed in terms of paths connecting $i$ and $j$, as depicted in FIG.\ref{takikawa7}.
On every bond in these paths, a $Z_2$ gauge field must be formed. 
For this reason, in each perturbation terms of
eq.(\ref{nthorder}), spin operators on all sites on the paths except the sites $i$ and $j$ should be expressed in terms of
gauge Majorana fields only, by using the relations, 
$ S^x_\ell = \frac{\rm i}{2}b^{x}_\ell c_\ell=-\frac{\rm i}{2}b^y_\ell b^z_\ell$, 
$S^y_\ell=\frac{\rm i}{2}b^{y}_\ell c_\ell=-\frac{\rm i}{2}b^z_\ell b^x_\ell$, and
$S^z_\ell=\frac{\rm i}{2}b^{z}_\ell c_\ell=-\frac{\rm i}{2}b^x_\ell b^y_\ell$.
Then, the lowest order terms which do not vanish in the vortex-free state are the third order perturbation terms, the explicit form of which  is given by,
\begin{eqnarray}
\mathcal{F}^{(3,3)}(\sigma_{i},\sigma_{j}\cdots)
 &:=& \Pi_{0}
 \Bigg[\sum_{\substack{  \Braket{ij}_{\alpha} \\ \beta,\gamma \neq \alpha}} {\sigma}_i^{\beta}{\sigma}_j^{\gamma}+{\sigma}_i^{\gamma}{\sigma}_j^{\beta}  \Biggr]^{3} 
\Pi_{0}\nn\\
&=&2\sum_{\alpha} \sum_{ \Braket{ij}_{\alpha} } (-c_{i}c_{j}b^{\gamma}_{j}b^{\beta}_{i}) \times
(-b^{\beta}_{k}b^{\alpha}_{k}b^{\alpha}_{l}b^{\gamma}_{l})\nn\\
&\times& (-b^{\gamma}_{s}b^{\beta}_{s}b^{\beta}_{t}b^{\alpha}_{t})\nn\\
&=&2\sum_{\alpha} \sum_{ \Braket{ij}_{\alpha} } {\rm i}c_{i}c_{j}\hat{u}^{\beta}_{ik} \hat{u}^{\alpha}_{kl} \hat{u}^{\gamma}_{ls}\hat{u}^{\beta}_{st}\hat{u}^{\alpha}_{jt}\nn\\
&=&2\sum_{\alpha} \sum_{ \Braket{ij}_{\alpha} } {\rm i}c_{i}c_{j}.
\end{eqnarray}
As a result, we obtain,
\begin{eqnarray}
\mathcal{H}^{(3)}_{eff} =-\frac{2\Gamma^{3}}{4^{3}}\sum_{\alpha} \sum_{\Braket{ij}_{\alpha} } {\rm i}c_{i}c_{j}, \label{eq:thirdorder}
\end{eqnarray}
which is a nearest neighbor hopping term.
This term gives only the change of the band width of itinerant Majorana fermions of the original Kitaev model
as $K \rightarrow K+\frac{2\Gamma^{3}}{4^{3}}$.
The above analysis for the non-vanishing lowest order term can be straightforwardly generalized to higher order terms.
It is found that, in all orders, any non-vanishing terms of eq.(\ref{nthorder}) give only nearest-neighbor hopping terms of itinerant Majorana fermions in the vortex-free state.
Therefore, we can conclude that the $\Gamma$ term and the Heisenberg term do not alter qualitative features of the Majorana band structure, 
as long as the Kitaev spn liquid state is not destabilized. 

\section{Summary}
In this paper, we investigated effects of non-Kitaev interactions, i.e. the Heisenberg exchange interaction, and the symmetric off-diagonal exchange interactions, the $\Gamma$ term and the $\Gamma'$ term, on the Kitaev spin liquid state by exploiting perturbative expansions 
around the vortex-free spin liquid state.
We demonstrated that the Heisenberg term and the $\Gamma$ term change the band width of itinerant Majorana fermions only, and
do not alter qualitative features of the spin liquid,
provided that the Kitaev spin liquid state is not destabilized.
On the other hand, it is found that
 the  $\Gamma'$ term drastically affects the band structure of Majorana fermions changing the number of the Dirac points,
and inducing topological phase transitions with the change of the Chern number from $\pm 1$ to $\pm 2$ under an applied magnetic field. 
The phase with the Chern number equal to $\pm 2$ is the Abelian topological phase, and can be detected by the measurement of the quantized thermal Hall conductivity.
Since the magnitude of $\Gamma'$ is quite sensitive to trigonal distortion of the crystal structure, 
it may be possible to realize the topological phase transition by applying strain on a sample.
 It is an interesting future issue to explore for the Abelian topological phase in candidate materials of Kitaev magnets.

Finally, we comment on the relation between our study and the recent related paper ref. \cite{Zhang}.
In ref. \cite{Zhang}, the realization of the Abelian topological phases in a Kitaev model with four-spin interaction terms are considered.
Some of the four-spin interaction terms considered in ref.\cite{Zhang}, which induce the transition to the Abelian topological phases, have the same form as the second-order perturbation term $\mathcal{H}^{(2)}_{\Gamma',eff}$ obtained in Sec.II in our paper.
Thus, our results provide a microscopic origin of the model considered in ref.\cite{Zhang}. 

\begin{acknowledgments}
The authors are grateful to Y. Kasahara and Y. Matsuda for valuable discussions. 
 This work was supported by the Grant-in-Aids for Scientific
Research from MEXT of Japan [Grants No. 17K05517, and KAKENHI on Innovative Areas ``Topological Materials Science'' [No.~JP15H05852]  and "J-Physics" [No.~JP18H04318], and JST CREST Grant Number JPMJCR19T5, Japan. 
\end{acknowledgments}

\bibliography{takikawa-ref}

\end{document}